\documentstyle[aps]{revtex}
\newcommand{\spmine}{1.10}
\newcommand{\myspace}{\edef\baselinestretch{\spmine}\Large\normalsize}
\begin{document}
\myspace
%
\title{The Phase Diagram of Correlated Electrons in a Lattice of Berry 
Molecules}
\vspace{10mm}
\author{Giuseppe Santoro, Marco Airoldi, Nicola Manini, Erio Tosatti}
\address{International School for Advanced Studies, Via Beirut 4, 34014 
Trieste, Italy}
\author{Alberto Parola}
\address{Istituto di Fisica, Via Lucini 3, Como, Italy}
%
\maketitle
\vspace{10mm}
\begin{abstract}
A model for correlated electrons in a lattice with local 
additional spin--1 degrees of freedom inducing constrained hopping, 
is studied both in the low density limit and at quarter filling.
We show that in both $1D$ and $2D$ two particles form a bound state even in 
presence of a repulsive $U<U_c$. 
A picture of a dilute Bose gas, leading to off-diagonal long range order (LRO)
in $2D$ (quasi-LRO in $1D$), is supported by quantitative calculations in 1D
which allow for a determination of the phase diagram.
\end{abstract}
%
%
\newpage

Recent work on models with correlated electronic hopping, involving 
density-dependent hopping matrix elements, has pointed out that 
superconductivity can arise by purely off-diagonal processes in spite
of repulsive short-range interactions.\cite{Cor_Hop}

Correlated electronic hopping arises also when additional degrees of 
freedom participate in the ``local physics'' at each site, inducing, 
for instance, {\em occupation-dependent constraints\/}. 
One clear example of the latter situation is realized in molecular crystals, 
where the additional degrees of freedom arise because of the coupling to
intra-molecular phonons.\cite{AMT,MTD}
The basic physics is that, under appropriate conditions, molecules with an 
{\em odd\/} number of electrons have an associated Berry phase and 
an orbital degeneracy, due to a dynamical Jahn-Teller effect, those with
even number have none.\cite{AMT}
The general question is whether such mechanisms will also lead to 
superconductivity even in the presence of repulsive diagonal interactions.

The simplest model capturing some of the physics of electronic hopping
in presence of additional local degrees of freedom is a Hubbard-type model 
with an extra {\em spin--1\/} at each site.\cite{MTD}
The constraint is that the allowed spin--1 states depend on the electron 
occupancy:\cite{MTD}
an empty or doubly occupied site must have $S^z=0$, whereas a 
site occupied by a single electron (either up or down) has an additional
twofold (orbital) degeneracy, represented by $S^z=\pm 1$, 
\begin{equation} \label{CONSTRAINT:eqn}
n_r=0,2 \, \rightarrow \, S^z_r=0 \;, \hspace{5mm}
n_r=1   \, \rightarrow \, S^z_r=\pm 1 \;. 
\end{equation}
The hamiltonian $H$ is written as follows:
\begin{equation} \label{MODEL:eqn}
H \,=\, - \frac{t}{2} \sum_{<rr'>}\sum_{\sigma} 
\; (c^{\dagger}_{r\sigma} c_{r'\sigma} + H.c.) \, (S^+_r S^-_{r'} + H.c.) 
\,+\, U \sum_r n_{r\uparrow} n_{r\downarrow} \;,
\end{equation}
where the $S^{\pm}_r$'s are spin--1 ladder operators at each site, and the 
remaining notation is completely standard.
It is worth stressing that while $H$, and in particular the hopping term, 
{\em conserves the constraint\/} in Eq.\ \ref{CONSTRAINT:eqn}, the model 
is still highly non-trivial, even for $U=0$.

In this paper we will present some results concerning the few 
particle problem, the low density region, and the quarter-filling case, 
which shed light on the basic features of the phase diagram of the model.
We will show that in both $1D$ and $2D$ the additional spin--1 degrees of
freedom lead to a two-electron bound state even in presence of a repulsive
$U<U_c$. Moreover, the model does not show tendency to phase separation and 
a picture of a dilute Bose gas, leading to off-diagonal quasi-long-range-order 
(LRO) in $1D$ (genuine LRO in $2D$) is supported at small density by 
analytical calculations. At quarter-filling, a clear superconducting regime
is found to survive up to a positive $U\sim t$.

Consider the two-particle problem first.
A state in the two-particle Hilbert space with total z-component of the spin 
$M^z_{TOT}=0$ (for both the electron spin and the spin--1 states) can be 
written as
$|\Psi\rangle = \sum_{r,r'} 
[ \psi_{+-}(r,r')  S^{+}_r S^{-}_{r'} + 
  \psi_{-+}(r,r')  S^{-}_r S^{+}_{r'}  ] 
c^{\dagger}_{r\uparrow} c^{\dagger}_{r'\downarrow} |0\rangle$, 
%
%
where the vacuum $|0\rangle$ is the state without fermions and with $S^z_r=0$
at each site. 
In writing $|\Psi\rangle$ we have taken into account the two possibilities of 
associating a $S^z=\pm 1$ spin state to the up and down electrons: 
$\psi_{+-}$ is the amplitude for having $S^z=+1$ associated to the 
$\uparrow$-electron (and $S^z=-1$ to the $\downarrow$-electron), while
$\psi_{-+}$ is the amplitude for other possible choice.
The Schr\"odinger equation for $\psi_{+-}(r,r')$ is easily shown to be
\begin{eqnarray} \label{Sch:eqn}
E \psi_{+-}(r,r') &\,=\,& -t\, \sum_{a} 
\, [ \, \psi_{+-}(r+a,r') + \psi_{+-}(r,r'+a) \, ]
\,+\, U \, \delta_{r,r'} \, \psi_{+-}(r,r') \nonumber \\
&& \hspace{04mm} 
-t\, (\sum_{a} \delta_{r+a,r'}) \; 
[ \, \psi_{-+}(r,r) + \psi_{-+}(r',r') \, ] \;,
\end{eqnarray}
where $a$ denotes a nearest neighbor vector ($a=\pm 1$ in 1D). 
A similar equation is obtained for $\psi_{-+}(r,r')$ by just exchanging 
$\psi_{-+}$ and $\psi_{+-}$ everywhere.
The last term in Eq.\ \ref{Sch:eqn} is crucial to the whole story, and deserves
a few comments. 
When the two electrons are far enough in the otherwise empty lattice,
the hamiltonian $H$ simply allows the hopping to a nearest neighbor site of 
the ``composite'' object formed by an electron and the associated spin--1 
state (first term in Eq.\ \ref{Sch:eqn}).
Things are more subtle when two electrons come to the same site $r$. 
In such a case, from a doubly occupied site with $S^z_r=0$ one can reach, 
upon hopping, two possible final states: either each electron keeps its own
spin--1 state or the spin--1 states associated to the two electrons are 
exchanged. It is precisely this second possibility of exchanging spin--1 states
that is responsible for the presence of $\psi_{-+}$ in the equation for 
$\psi_{+-}$ and vice-versa (last term in Eq. \ref{Sch:eqn}).

The Schr\"odinger equation is easily solved in momentum space, where it reduces
to a $2\times 2$ matrix problem. 
For total momentum $P=0$, the set of solutions among which the ground state is 
found satisfies the equation
\begin{equation} \label{E:eqn}
\frac{1}{L^D} \sum_{k} \frac{1}{E-2\epsilon_k} = \frac{2}{E+U} \;,
\end{equation}
where $E$ is the energy eigenvalue and $\epsilon_k$ is the tight-binding 
dispersion of the free-electron problem 
($\epsilon_k=-2t\sum_{\alpha}\cos{k_{\alpha}}$). 
In the ordinary Hubbard case, the right hand side 
of Eq.\ \ref{E:eqn} would simply read $1/U$.\cite{creta}
A graphical analysis of Eq.\ \ref{E:eqn} readily shows that a bound state 
solution is present even for $U>0$ up to $U_c=4Dt$ in $D\leq 2$. 
In $D\geq 3$ a finite attractive $U$ is needed to produce a bound state.

The bound state solution can be worked out analytically in 1D. For
general values of the total momentum $P$ and for $U\le U_c=4t\,\cos(P/2)$, 
the ground state energy is given by
\begin{equation} \label{E_BS:eqn}
E_P \,=\, -4t \, \cos{(P/2)} \; (Z_P^2 + 1)/(2Z_P) \;,
\end{equation}
%
%
with $Z_P=\{ -(U/4t) + [ (U/4t)^2 + 3\cos^2{(P/2)} ]^{1/2} \}/\cos{(P/2)}$,
and the corresponding ground state wave-function is
$\psi_{+-}=\psi_{-+}=e^{iP(r+r')/2}[e^{-\kappa |r-r'|}-(1/2)\delta_{r,r'}]$ 
%
%
with $\kappa=\ln{Z_P}$. At larger values of $U$, no bound state is
present and the energy spectrum is continuous in the infinite lattice.
There is still, however, an antibound state. The two-particle solution cannot
be generalized to an arbitrary number of particles by Bethe Ansatz because the
corresponding scattering matrix does not satisfy the Yang-Baxter relations.

The form of the ground-state wave function naturally provides a picture of 
bound pairs approximately localized on adjacent lattice sites, thereby forming 
{\sl dimers}. Remarkably, the rather strong attraction responsible for this 
binding is {\sl generated by the kinetic term alone} via the presence of the 
additional degrees of freedom.
The critical value of the Hubbard repulsion $U_c=4t$ is considerably larger 
than the ground state binding energy at $U=0$: 
$E_b=4t\,(Z_0-1)^2/(2Z_0)\approx 0.618\,t$ 
showing that this kind of pairing mechanism is rather insensitive to the
presence of on site Coulomb repulsion. 
This is explained by its ``kinetic'' origin which delocalizes the pair on 
neighboring sites. The same feature is also present in 2D where 
$U_c=8t$: The enhancement is due to the the larger coordination of the 
2D lattice which provides an even more efficient delocalization of the 
electron pair.

This interpretation of the two-particle ground state in terms of dimers 
leads to a simple picture, both in $1D$ and $2D$, of the low density 
region of the model (\ref{MODEL:eqn}), a picture which we will explicitly 
verify in $1D$:
For $U<U_c$, at low densities, the system behaves as a weakly interacting, 
dilute gas of dimers, which follow boson statistics and may be thought of 
as bosons with an extended core.
In $1D$, we can estimate the dimer effective mass from our two-particle 
calculation, Eq.\ \ref{E_BS:eqn}, which at small momentum $P$ gives 
$E_P \sim E_0 + t_{\rm eff} P^2$ with 
$t_{\rm eff}=t\left[3+(U/4t)^2\right]^{-1/2}$. 
The effective mass has a smooth dependence on $U$ varying between $\sqrt{3}$ 
at $U=0$ and $2$ at $U=4t$.
Dilute dimers are therefore rather mobile and a superfluid ground state
must be expected at zero temperature. In $1D$, of course, off-diagonal LRO 
cannot occur and only a long range power-law decay of the dimer density 
matrix is possible, while in $2D$ a genuine Bose condensate will form. 
In terms of the original electrons this implies a standard strong coupling 
BCS superconducting ground state with localized Cooper pairs. 
A similar scenario has also been proposed in the framework of the one 
dimensional $t-J$ model where bound pairs are formed at low 
density \cite{zurigo} and $2<J/t<2.95$. In that case, however, the model is 
unstable to phase separation, which in fact occurs massively at larger values 
of $J/t$.

To check whether the superfluid dimer picture is correct and that no phase 
separation occurs at low density, we have carried out a numerical 
investigation of the four particle problem by exact diagonalization 
of the model (\ref{MODEL:eqn}) in lattices with up to 24 sites.
The results for the ground state energy and for the density--density
correlation function at $U=0$ are reported in Fig.\ 1.
The four electrons are sharply localized in two pairs placed at the maximum 
separation. 
The origin of this repulsion is purely kinetic and is also reproduced by 
the ground state wave function of two hard core bosons in $1D$,
$\psi(R) \propto |\sin{(\pi R/L)}|$. 
The scaling of the energy of four electrons clearly shows that the limiting 
value is just twice the pair energy $E_0=-8 t /\sqrt{3}$ 
with $1/L^2$ corrections. 
Free ``hard core bosons'' with the appropriate effective
dispersion $t_{\rm eff}=t \left [3 + (U/4t)^2\right ]^{-1/2}$, give
the asymptotic result $L^2(E -2\,E_0) \to {2\pi^2\,t_{\rm eff}}$ marked 
with a full circle in Fig.\ 1, and clearly compatible with the data. 

A further step in the quantitative characterization of the low
density region of the model can be made by using standard 
field theoretical methods for $1D$ systems.\cite{oned}
The analysis of the two and four-particle problem shows that for
$U<U_c$ a spin gap is present in the excitation spectrum of the model while 
the charge (superfluid) degrees of freedom remain gapless, giving origin to 
the power-law scaling of the ground state energy shown in Fig.\ 1. 
Above $U_c$ a two-electron bound state does not form and the triplet 
spectrum looks gapless. This is actually not the case at finite density, 
as can be inferred from the analysis of the $U\to \infty$ limit of the 
model.\cite{santorini}
The strong coupling expansion of the model in Eq.\ \ref{MODEL:eqn} can be 
obtained closely following the analogous calculation in the 
Hubbard model.\cite{hubtj} The resulting hamiltonian, to $O(t^2/U)$,
is defined in the subspace of singly occupied sites where electrons
are characterized by {\it two} internal spin-$1/2$ degrees of freedom: 
The usual spin ($\sigma$) and a pseudospin ($\tau$). 
The latter takes into account the two possible states of the original
spin--1 allowed by the constraint (\ref{CONSTRAINT:eqn}).
The resulting hamiltonian contains a free hopping term
(with coupling $-t$) and a spin (and pseudospin) dependent 
part (with coupling $\propto t^2/U$). In 1D, analogously to the 
Hubbard model case,\cite{ogata} the wave function factorizes, at $U\to\infty$, 
for all densities: 
The position of the electrons is determined by a spinless fermion wave function 
while the spin (and pseudospin) ordering is governed by the effective 
spin hamiltonian
\begin{equation} \label{spinham}
H_{ST}= - J_{\rm eff} \sum_{<i,j>}
\left [ \vec\sigma_i\cdot \vec\sigma_j - 1/4 \right ]
\left [ \vec\tau_i  \cdot \vec\tau_j   - 1/4 \right ] \;,
\end{equation}
with $J_{\rm eff}$ depending on the electron density $\rho=N/L$ as 
$J_{\rm eff} = (8t^2/U) \rho [1-\sin(2\pi\rho)/2\pi\rho]$. 
Numerical as well as variational calculations have shown
that the hamiltonian in Eq.\ \ref{spinham} is characterized by a 
{\em spin gap\/} of order $J_{\rm eff}$.\cite{santorini} 
Fig.\ 2 (inset) shows the finite size spin gap at filling $\rho=1/2$ 
for several values of $U$. A finite spin gap is quite clear from the 
$L\to \infty$ extrapolations. We also find that the gap persists to large 
$U$ and correctly scales with $1/U$. 
Similar results are also obtained at half-filling, where
Eq.\ \ref{spinham} is the large $U$ effective hamiltonian.
We conclude that our model (Eq.\ \ref{MODEL:eqn}), at arbitrary (finite) 
density, has a spin gap all the way to $U\to\infty$. 
In the zero--density limit $\rho\to 0$, however, the spin gap vanishes 
very fast, like $\rho^3$, whence the gapless appearance of our previous 
two-particle results.

The spin (and pseudospin) degrees of freedom are gapped at all $U$ and 
only the charge degrees of freedom remain gapless. 
The universality class of the hamiltonian in Eq.\ \ref{MODEL:eqn} is then 
given by the Luther--Emery model, \cite{solyom} and the long range decay 
of the correlation functions is characterized by power-laws which can be 
obtained by the knowledge of the single exponent $K_{\rho}$. \cite{oned} 
This exponent is also related by the Haldane--Schulz equation to the 
compressibility of the model:
\begin{equation} \label{hs}
L{\partial^2 E\over \partial N^2}={\pi\over 4}{u_{\rho}\over K_{\rho}} \;.
\end{equation}
The factor $4$ at right hand side takes into account the presence 
of an additional (pseudospin) degree of freedom which can be attributed
to the electron in the present model, due to the additional 
degeneracy $S_z=\pm 1$ in Eq.\ \ref{CONSTRAINT:eqn}. 
Correspondingly, $N$ includes both spin and pseudospin degrees of freedom. 
As usual, $u_{\rho}$ represents the charge velocity, i.e., the velocity 
associated with the branch of gapless excitations. 
In order to use Eq.\ \ref{hs} for obtaining the low density limit 
of $K_{\rho}$, we should calculate both the ground state
energy and the charge velocity at small but finite density. However, 
assuming a smooth zero density limit, it is possible to get the same 
information from the two and four-particle solution. In fact, at
{\sl fixed N}, and $L\to \infty$, the predicted scaling of the energy is
$E(N,L)=N\epsilon_0 + \alpha N^3/L^2 -\beta N/L^2+O(L^{-3})$.
Analogously, the large $L$ (i.e., low density) limit of the charge 
velocity should scale as $u_{\rho}=\gamma N/L$, 
where $\alpha$, $\beta$ and $\gamma$ are model-dependent constants. 
By means of these asymptotic expressions together with Eq.\ \ref{hs}, 
we get the formal result for $K_{\rho}=\pi\gamma /24\alpha$.
The central charge $c$ \cite{korepin} of the model can be also expressed in 
terms of these constants, $c=6\beta/\pi\gamma$. This will provide a 
consistency check. 
At $U>U_c=4t$, the two-particle analytical result is sufficient to
extract the values of the required constants, which turn out to be independent
of $U$: $\alpha=\pi^2/24$, $\beta=\pi^2/6$ and $\gamma=\pi$. 
This gives $c=1$, as expected from the universality class of the model, 
and $K_{\rho}=1$. 
In order to extract the correct large $L$ scaling of the energy in the
regime $U<U_c$, the four-particle numerical result is required. 
By using the estimate of Fig.\ 1, the parameters governing the asymptotic 
low density behavior become $\alpha =t_{\rm eff} \pi^2/ 192$, 
$\beta=t_{\rm eff} \pi^2/12$ and $\gamma= t_{\rm eff} \pi/2$, which give
$c=1$ and $K_{\rho}=4$ independent of $U<4t$. 

The exponents which characterize the decay of the electronic correlation 
functions are obtained in terms of the parameter $K_{\rho}$ by bosonization 
under the assumption that only the charge sector remains gapless.\cite{solyom} 
In particular, the density response function at $2 k_F$ decays as 
$x^{-K_{\rho}/2}$ while the $4 k_F$ component behaves as $x^{-2 K_{\rho}}$. 
In the derivation of these power-laws, we must recall that eight bosonic fields
contribute to the physical density operators, and this reduces the exponents 
by a factor 2 with respect to the usual Hubbard model case where only four 
fields occur (right and left movers with two possible values of the spin). 
At the Luther--Emery fixed point also the superconductive correlations have 
power-law behavior.
By examining the corresponding operator, we get the asymptotic behavior of 
the s--wave superconductive correlation as $\Delta_s(x)\sim x^{-2/K_{\rho}}$. 
At low densities and for $U<4t$ we have $K_{\rho}=4$: the most relevant 
correlations are the superconductive ones, decaying as $1/\sqrt{x}$. 
Remarkably, the $1/\sqrt{x}$ behavior is identical with that of a dilute hard 
core boson gas, as expected from the physical picture previously discussed. 
Instead, for $U>4t$ and $\rho\to 0$, we have $K_{\rho}=1$:
the correlation with the slowest decay is the density response function at 
$2 k_F$, with an inverse square root behavior, implying a divergence in the 
CDW susceptibility. 
Strictly at $U=\infty$ the $2k_F$ response function has vanishing amplitude 
(like in the Hubbard model) and therefore the corresponding singularity is 
absent in the exact solution previously discussed: we expect that, at 
finite $U$, this additional feature may be detected.

We have also determined $K_{\rho}$ numerically from the charge velocity
$u_{\rho}$ and from the Drude peak $D=u_{\rho}K_{\rho}/(2\pi)$.\cite{oned} 
Fig.\ 2 shows the results obtained at quarter-filling ($\rho=1/2$). 
For $U/t<1$ the $L\to \infty$ extrapolation clearly gives $K_{\rho}>2$, i.e., 
predominantly superconducting correlations.
On the contrary, for $U/t>1$ we have $K_{\rho}<2$, i.e., predominant CDW 
correlations. The crossover between the two regimes is found to take place
close to $U\approx t$ at this density. Exactly at half-filling ($\rho=1$) 
a transition is found at $U_c\approx 0$ between a superconductor ($U<U_c$)
and a Mott insulator fully gapped in both spin and charge 
sector.\cite{santorini}

The physics of the model in one dimension is now rather clear. 
A transition occurs at $U_c(\rho)$ (with $U_c(0)=4t$, $U_c(1/2)\approx t$,
and $U_c(1)\approx 0$) between a superconductor ($U<U_c$) and a sliding 
CDW conductor ($\rho\ne 1$) or Mott insulator ($\rho=1$). 
The physics at small $U$, leading to superconducting correlations, is rather
robust and can be generalized to $2D$ where true LRO will occur at zero 
temperature.
On the other hand, the CDW state, perhaps akin to a spin Peierls state, 
should get weaker and disappear in higher dimensions. The large $U$ spin gap
is also, probably, a $1D$ peculiarity. 

By extrapolating these results, we can surmise that a correlated hopping model
such as Eq.\ 2 can be expected to give rise to superconductivity in $3D$. 
A pairing mechanisms based on correlated hopping is not easily 
destroyed by a repulsive $U$, it is more effective at low carrier density, 
and is apparently immune from the polaron self-trapping, which depresses 
$T_c$ in strongly coupled electron-phonon systems.
More work is now being devoted to investigate the relevance of this type
of model to molecular superconductors such as organic $1D$ metals, Chevrel
phases, and metal fullerides.

It is a pleasure to thank M. Fabrizio, S. Sorella, and S. Doniach 
for many enlightening discussions. The sponsorship of NATO, through CRG 920828,
is gratefully acknowledged. 


\newpage
\begin{center}
{\bf Figure Captions}
\end{center}

\begin{description}

\item[Figure 1] Size scaling of the ground state energy of the hamiltonian 
in Eq.\ 2 with $U=0$ and four electrons. 
$\Delta E=E_L(4)-2E_{\infty}(2)$ where $E_L(N)$ is the ground state energy
of $N$ electrons in a $L$-site ring in units of $t$. 
The full circle is the exact result for two hard-core bosons in the infinite 
lattice with $t_{\rm eff}=t/\sqrt{3}$, fully compatible with the $L\to \infty$
extrapolation of the data. 
Inset: the density-density correlation function for four electrons in a 
$L=20$-site ring at $U=0$.

\item[Figure 2] 
Size scaling for the exponent $K_{\rho}$, obtained from finite size values
of the Drude peak $D$ and charge velocity $u_{\rho}$ at filling 
$\rho=1/2$, for $U/t=0,1,2$, and $4$. 
Inset: the spin gap for the same parameters. The continuous lines represent
fits of the form $\Delta E=a+b/L+c/L^2$ and $K_{\rho}=d+e/L^2$. 
Note the transition from superconductive ($K_{\rho}>2$) to CDW correlations
($K_{\rho}<2$) for $U/t\approx 1$. The spin gap is always finite.

\end{description}

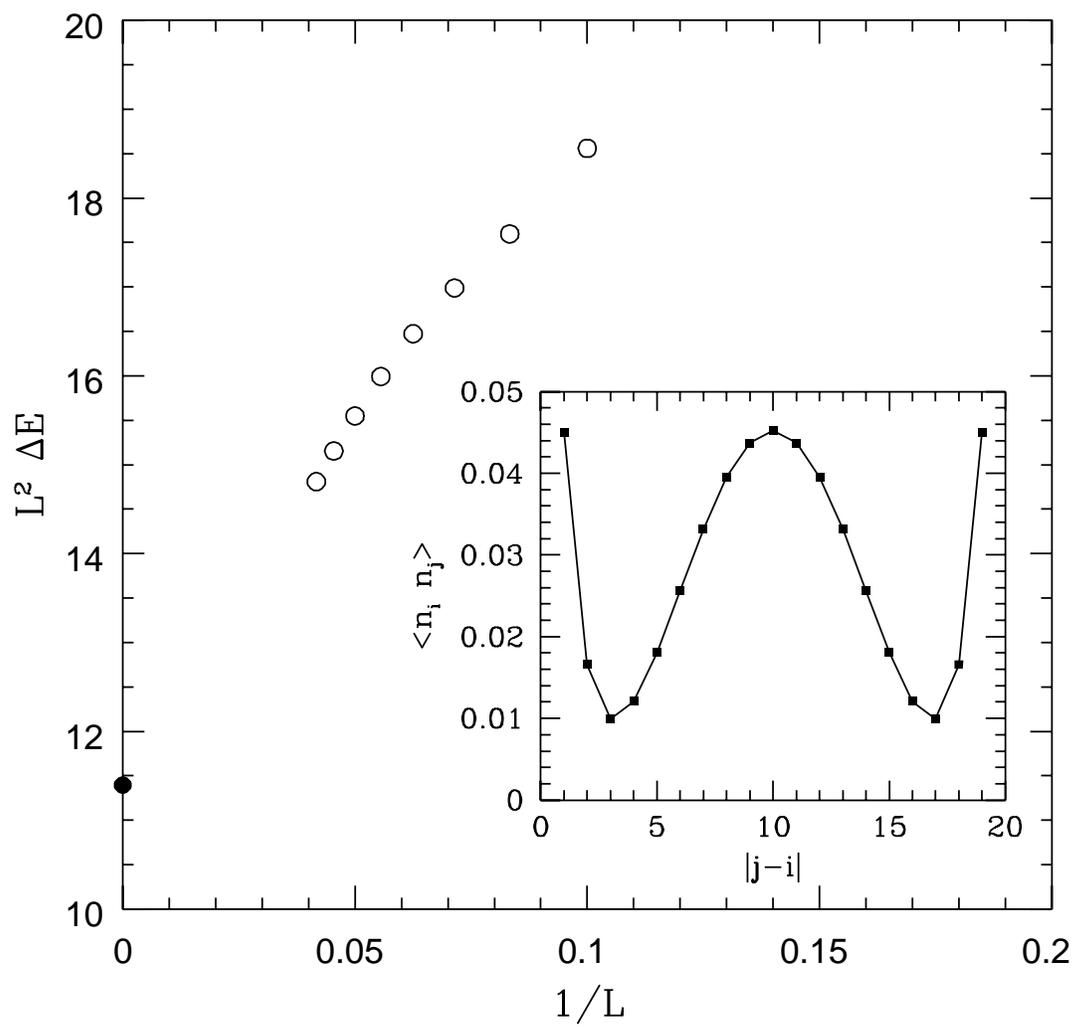
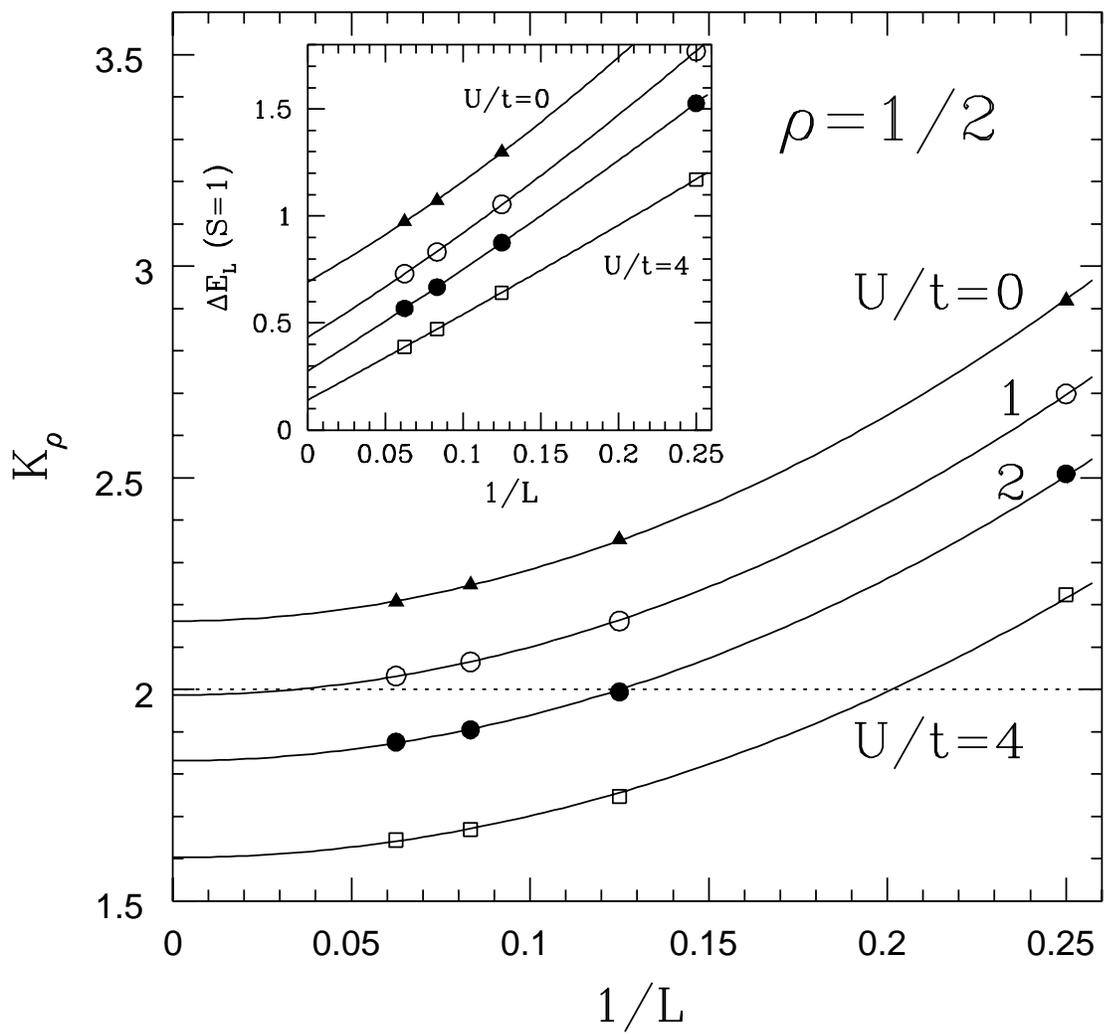
\end{document}